\documentclass[12pt,centertags,english]{article}
\newcounter{withlatexpictures}
\usepackage{babel}
\usepackage{a4}
\usepackage{amsmath}
\usepackage{amssymb}
\ifnum\thewithlatexpictures=1
  \usepackage{epic}
\fi
\usepackage{epsfig}
\usepackage{subfigure}
\usepackage{fancyheadings}
\makeatletter\ifnum\@ptsize=2
\fi\makeatother%
\makeatletter
\renewcommand{\@makecaption}[2]{
  \vskip\abovecaptionskip%
  \sbox\@tempboxa{#1: \emph{#2}}%
  \ifdim \wd\@tempboxa >\hsize%
    #1: \emph{#2}\par%
  \else%
    \global \@minipagefalse%
    \hb@xt@\hsize{\hfil\box\@tempboxa\hfil}%
  \fi%
  \vskip\belowcaptionskip%
}
\makeatother
\newcommand{\refeq}[1]{(\ref{eq:#1})}
\newcommand{\abs}[1]{\ensuremath{\lvert #1\rvert}}
\newcommand{\amplitude}[1]{\ensuremath{\mathcal{A}_{#1}(\xi)}}
\newcommand{\anti}[1]{\ensuremath{\bar{#1}}}
\newcommand{\carlson}[4]{\ensuremath{%
  \mathop{{}R_{#1}}({\textstyle#2,#3,#4})}}
\newcommand{\chpolynom}[2]{\ensuremath{%
  \mathop{{}T_{#1}}({\textstyle#2})}}
\newcommand{\defeq}{\ensuremath{:=}}
\newcommand{\diffone}[1][]{\ensuremath{\mathop{\mathrm{d}#1}\nolimits}}
\newlength{\dcl}
\newcommand{\dirac}[2][0pt]{\ensuremath{\text{\settowidth{\dcl}{$#2$}$#2
  $\hspace{-0.5\dcl}\makebox[0pt][c]{\hspace{#1}$/$}\hspace{0.5\dcl}}}}
\newcommand{\efp}[4][]{\ensuremath{%
  \mathop{{}#2_{#3}^{#1}}({\textstyle#4})}}
\newcommand{\gbmoment}[3]{\ensuremath{%
  \mathop{{}G_{#1}^{#2}}({\textstyle#3})}}
\newcommand{\gbpolynom}[3]{\ensuremath{%
  \mathop{{}C_{#1}^{(#2)}}({\textstyle#3})}}

\newcommand{\intoned}[4][]{\ensuremath{%
  \int_{#3}^{#4}#1\diffone[#2]}}
\newcommand{\invkernel}[4]{\ensuremath{%
  \mathop{{}\mathcal{K}^{-1}_{#1}({\textstyle#2,#3;#4})}}}
\newcommand{\kernel}[4]{\ensuremath{%
  \mathop{{}\mathcal{K}_{#1}({\textstyle#2,#3;#4})}}}
\newcommand{\landauO}[1]{\ensuremath{%
  \mathop{{}\mathcal{O}({\textstyle#1})}}}
\newcommand{\mcomma}{\ensuremath{\;,}}
\newcommand{\mmoment}[2]{\ensuremath{\mathop{{}M_{#1}^{#2}}{}}}
\newcommand{\mperiod}{\ensuremath{\;.}}

\newcommand{\ofh}[4][]{\ensuremath{%
  \mathop{{}H_{#2}^{#1}}({\textstyle#3,#4})}}
\newcommand{\opdelta}[2][]{\ensuremath{%
  \mathop{{}\delta^{#1}}({\textstyle#2})}}

\newcommand{\opGamma}[1]{\ensuremath{%
  \mathop{{}\Gamma}({\textstyle#1})}}
\newcommand{\optheta}[1]{\ensuremath{%
  \mathop{{}\theta}({\textstyle#1})}}
\DeclareMathOperator{\disc}{disc}
\DeclareMathOperator{\imagpart}{Im}
\DeclareMathOperator{\Log}{Log}
\DeclareMathOperator{\realpart}{Re}

\usepackage{hyperref}
\ifnum\thewithlatexpictures=1
  \usepackage{eepic}
\fi
\begin{document}
%
%
\allowdisplaybreaks
%
%
\begin{titlepage}
  \title{
    \begin{flushright}\normalsize
      DO-TH 00/08\\
      April 2000
    \end{flushright}
    \begin{bfseries}
      Off-forward parton distributions and\\ Shuvaev's transformations
    \end{bfseries}
  }
  \author{
    Jens~D.~Noritzsch%
    \thanks{Email address: \texttt{jensn@hal1.physik.uni-dortmund.de}}\\
    \begin{itshape}
      Institut f\"{u}r Physik, Universit\"{a}t Dortmund,
      D-44221 Dortmund, Germany
    \end{itshape}
  }
  \date{}
\end{titlepage}
\maketitle
%
%
\begin{abstract}
  We review Shuvaev's transformations that relate off-forward parton
  distributions (OFPDs) to the so-called effective forward parton
  distributions (EFPDs). The latter evolve like conventional forward
  partons.  We express nonforward amplitudes, depending on OFPDs,
  directly in terms of EFPDs and construct a model for the EFPDs, which
  allows us to consistently express them in terms of the conventional
  forward parton distributions and nucleon form factors.  Our model is
  self-consistent for arbitrary $x$, $\xi$, $\mu$, and $t$.
\end{abstract}
%
%
\section{Introduction}
The treatment of nonforward high-energy processes, such as deeply
virtual Compton scattering (DVCS) and hard exclusive electroproduction
of vector mesons, in perturbative QCD gives rise to a new class of
parton distributions, the so-called skewed parton distributions
(SPDs)~\cite{Dittes:1988xz,Muller:1994fv,Ji:1997ek,Ji:1997nm,Ji:1998pc,%
Collins:1997fb,Radyushkin:1996ru,Radyushkin:1997ki} and double
distributions (DDs)~\cite{Radyushkin:1996nd,Radyushkin:1997ki},
generalizing the well-known conventional parton
distributions~\cite{Gluck:1998xa,Lai:1999wy,Martin:1999ww} and, at the
same time, the nucleon form factors. In the following we restrict
ourselves to the off-forward parton distributions (OFPDs) introduced by
Ji~\cite{Ji:1997ek,Ji:1997nm,Ji:1998pc}, which are equivalent to
nonforward~\cite{Radyushkin:1996ru,Radyushkin:1997ki} or
off-diagonal~\cite{Collins:1997fb} parton distributions.  Therefore, our
results can be easily generalized to nonforward and off-diagonal parton
distributions.

The off-forward parton distributions $\ofh{p}{x}{\xi,t,\mu}$, which
parametrize nonforward matrix elements of light cone bilocal operators
$\langle{}P'\lvert\mathcal{O}_{p}(-n/2,n/2)\rvert{}P\rangle\rvert_{n^{2}
=0}$, depend on the momentum fraction $x$ of the average nucleon
momentum $\bar{P}\defeq(P+P')/2$, which the initial state parton $p$
carries, on the ``skewedness''
$\xi=-n\cdot\Delta/2\mathop{n\cdot\bar{P}}$ with $\Delta\defeq{}P'-P$,
on the momentum transfer invariant $t\defeq\Delta^{2}$, and on the
renormalization scale $\mu$.  For vanishing $\xi$ and $t$, they are
identical to the usual forward parton distributions.  Detailed reviews
on off-forward parton distributions can be found, e.g., in
Refs.~\cite{Radyushkin:1997ki,Ji:1998pc}.

Recently, Shuvaev~\cite{Shuvaev:1999fm} demonstrated that the
off-forward parton distributions can be related (at least in leading
order) by simple transformations to so-called effective forward parton
distributions (EFPDs), the renormalization scale dependence of which is
governed by the conventional forward evolution equations.  These
relations have led to some progress in determining the shape of the
off-forward parton distributions for small values of
$\xi$~\cite{Shuvaev:1999ce}, since the EFPDs can be identified with the
usual partons for small values of $\xi$ and arbitrary scale $\mu$.

In the present paper, we express nonforward amplitudes directly in terms
of effective forward parton distributions. Furthermore, we define a
family of self-consistent models for EFPDs, in which the effective
forward parton distributions are obtained from the conventional forward
parton distributions and nucleon form factors at arbitrary scale $\mu$.

In the next section we briefly review the basic properties of the
off-forward parton distributions, and we define the effective forward
parton distributions.  In Sec.~\ref{s:invkernel}, we recalculate
Shuvaev's inverse transformations, which relate the EFPDs to the OFPDs,
and we derive their support in $x$, which is not identical to
$-1\leq{}x\leq1$ as for the conventional forward parton
distributions.\footnote{We use throughout parton distributions with both
signs of $x$, i.e., $\efp{q}{}{-x}=-\efp{\anti{q}}{}{x}$ and
$\efp{g}{}{-x}=-\efp{g}{}{x}$.}  In Sec.~\ref{s:kernel}, Shuvaev's
transformation is brought into a form that is convenient for a further
analytical and numerical treatment.  In Sec.~\ref{s:amplitude}, we
connect the effective forward parton distributions directly to
nonforward amplitudes, and we briefly discuss the reliability of simple
approximative formulas.  In Sec.~\ref{s:model}, we introduce our
model. Finally, in Sec.~\ref{s:summary}, we summarize our results, and
we draw the conclusions.
%
%
\section{Off-forward and effective parton distributions}
\label{s:ofpd}
The long-distance behavior of hard scattering processes, which is not
calculable in (QCD) perturbation theory, is factorized in matrix
elements of light-cone bilocal operators.  A Fourier transformation of
diagonal matrix elements results in the conventional quark and gluon
densities $q(x)$ and $g(x)$.  Analogously, the off-forward parton
distributions are defined by nonforward matrix elements:
\begin{subequations}\begin{align}
  &\left.\left\langle{}P',S'\left\lvert\anti{\psi}_{q}(-\tfrac{n}{2})
    \left.\dirac{n}\mathcal{G}\right.\psi_{q}(\tfrac{n}{2})\right
    \rvert{}P,S\right\rangle\right\rvert_{n^{2}=0}\notag\\
  &\qquad=\anti{U}(P',S')\left.\dirac{n}\right.U(P,S)\intoned[{e^{-ix(
    n\cdot\anti{P})}\ofh{q}{x}{\xi,t}}]{x}{-1}{+1}+\landauO{\Delta}
    \mcomma\\
  &\left.\left\langle{}P',S'\left\lvert{}F^{a}_{\mu\lambda}(-\tfrac{n}
    {2})\left.n^{\mu}n^{\nu}\mathcal{G}_{ab}\right.F^{b\lambda}{}_{\nu
    }(\tfrac{n}{2})\right\rvert{}P,S\right\rangle\right\rvert_{n^{2}=0
    }\notag\\
  &\qquad=\frac{1}{2}\left.\anti{U}(P',S')\left.\dirac{n}\right.U(P,S)
    \right.\left(n\cdot\anti{P}\right)\intoned[{e^{-ix(n\cdot\anti{P})
    }\ofh{g}{x}{\xi,t}}]{x}{-1}{+1}+\landauO{\Delta}\mcomma
\end{align}\end{subequations}
where $\mathcal{G}_{(ab)}$ is the Wilson gauge link.  Note that we have
an additional factor of $x$ in the definition of the gluon distribution
compared to the original definition of
Ji~\cite{Ji:1997ek,Ji:1997nm,Ji:1998pc},
$H_{g}^{}=x\mathop{H_{g}^{\text{Ji}}}$, which removes an ``artificial''
singularity for finite $\xi$~\cite{Radyushkin:1997ki}.  Due to time
reversal invariance and hermiticity, Ji's OFPDs are even functions of
$\xi$, so it is sufficient to treat only positive values of $\xi$.  The
different $\landauO{\Delta}$ contributions can be found, for example, in
Refs.~\cite{Radyushkin:1997ki,Ji:1998pc,Polyakov:1999gs}.  For vanishing
$\Delta$ the off-forward parton distributions reduce to the diagonal
partons:
\begin{subequations}\begin{align}
  \ofh{q}{x}{0,0}&=\efp{q}{}{x}\mcomma\\
  \ofh{g}{x}{0,0}&=x\efp{g}{}{x}\mperiod
\end{align}\end{subequations}
The renormalization of the defining operators leads to a scale
dependence of the off-forward parton distributions.  The evolution of
the OFPDs takes a simple form at the one-loop level for the Gegenbauer
moments,
\begin{subequations}\label{eq:gm}\begin{align}
  \gbmoment{n}{q}{\xi,t,\mu}&\defeq\frac{2^{n}[n!]^{2}}{(2n+1)!}\intoned
    [{\xi^{n}\gbpolynom{n}{3/2}{\tfrac{x}{\xi}}\ofh{q}{x}{\xi,t,
    \mu}}]{x}{-1}{+1}\mcomma\\
  \gbmoment{n}{g}{\xi,t,\mu}&\defeq\frac{3\cdot2^{n}(n-1)!n!}{(2n+1)!}
    \intoned[{\xi^{n-1}\gbpolynom{n-1}{5/2}{\tfrac{x}{\xi}}\ofh{
    g}{x,t,\mu}{\xi}}]{x}{-1}{+1}\mcomma
\end{align}\end{subequations}
since they evolve exactly as the Mellin moments in the diagonal
case~\cite{Radyushkin:1997ki}.  For example, for the Gegenbauer moments
of the nonsinglet off-forward quark distributions one has
\begin{equation}
  \gbmoment{n}{q,ns}{\xi,t,\mu}=\left(\frac{\alpha_{s}(\mu)}{\alpha_{s}(
    \mu_{0})}\right)^{\gamma_{0n}/2\beta_{0}}\gbmoment{n}{q,ns}{
    \xi,t,\mu_{0}}\mcomma
\end{equation}
where $\gamma_{0n}$ and $\beta_{0}$ are the leading coefficients of the
nonsinglet anomalous dimension and the beta function.  This fact allows
the definition of effective forward parton distributions, whose Mellin
moments equal the corresponding Gegenbauer moments of the
OFPDs~\cite{Shuvaev:1999fm}:
\begin{subequations}\label{eq:efpd}\begin{align}
  \intoned[{x^{n}\efp{q}{\xi,t}{x,\mu}}]{x}{-1}{+1}&=\gbmoment{n}{q}{\xi
    ,t,\mu}\mcomma\\
  \intoned[{x^{n}\efp{g}{\xi,t}{x,\mu}}]{x}{-1}{+1}&=\gbmoment{n}{g}{\xi
    ,t,\mu}\mperiod
\end{align}\end{subequations}
Their scale dependence is governed by the conventional evolution
equations, and they reduce to the diagonal quark and gluon densities for
$\xi,t\to0$.  The crucial point is that the effective and off-forward
parton distributions can be related to each other by Shuvaev's
transformations~\cite{Shuvaev:1999fm}.  As these transformations do not
depend on the momentum transfer invariant $t$ and the scale $\mu$, we
can safely skip them in the following.
%
%
\section{Shuvaev's inverse transformation}
\label{s:invkernel}
\begin{figure}
  \hspace{\stretch{2}}
  \subfigure[]{
    \ifnum\thewithlatexpictures=1
      \setlength{\unitlength}{1cm}
      \begin{picture}(6,4)
        \thinlines\footnotesize
          \put(0,0){\shortstack[l]{$\vphantom{\dfrac{1}{w}}w>1$\\$z_{\pm
          }=w\pm\sqrt{w^{2}-1}$}}
          \put(0,2){\vector(1,0){6}}
          \put(6,2.1){\makebox(0,0)[br]{$\realpart{}z$}}
          \put(3,0){\vector(0,1){4}}
          \put(2.9,4){\makebox(0,0)[tr]{$\imagpart{}z$}}
          \put(4,1.9){\line(0,1){0.2}}
          \put(4,2.1){\makebox(0,0)[bl]{$1$}}
          \put(3,2){\circle{1}}
          \put(3.5,1.9){\line(0,1){0.2}}
          \put(3.5,1.9){\makebox(0,0)[tl]{$z_{-}$}}
          \put(5,1.9){\line(0,1){0.2}}
          \put(5,1.9){\makebox(0,0)[tl]{$z_{+}$}}
        \Thicklines
          \put(3.5,2){\line(1,0){1.5}}
      \end{picture}
    \else
      \epsfig{figure=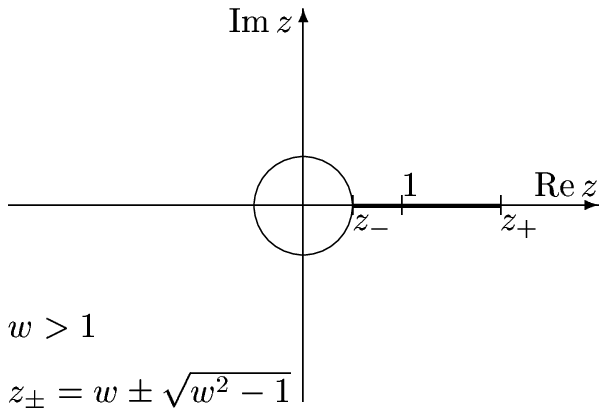}
    \fi
  }
  \hspace{\stretch{1}}
  \subfigure[]{
    \ifnum\thewithlatexpictures=1
      \setlength{\unitlength}{1cm}
      \begin{picture}(6,4)
        \thinlines\footnotesize
          \put(0,0){\shortstack[l]{$\dfrac{1}{w}>1$\\$z_{\pm}=w\pm i
            \sqrt{1-w^{2}}$}}
          \put(0,2){\vector(1,0){6}}
          \put(6,2.1){\makebox(0,0)[br]{$\realpart z$}}
          \put(3,0){\vector(0,1){4}}
          \put(2.9,4){\makebox(0,0)[tr]{$\imagpart z$}}
          \put(4,1.9){\line(0,1){0.2}}
          \put(4,2.1){\makebox(0,0)[bl]{$1$}}
          \put(3,2){\circle{2}}
          \put(3.7,2.7){\line(1,0){0.1}}
          \put(3.8,2.7){\makebox(0,0)[l]{$\vphantom{z_{-}}z_{+}$}}
          \put(3.7,1.3){\line(1,0){0.1}}
          \put(3.8,1.3){\makebox(0,0)[l]{$\vphantom{z^{+}}z_{-}$}}
        \Thicklines
          \put(3.7,2.7){\line(0,1){1.3}}
          \put(3.7,1.3){\line(0,-1){1.3}}
      \end{picture}
    \else
      \epsfig{figure=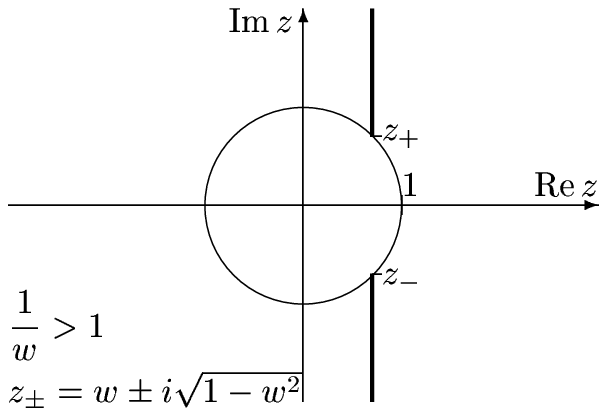}
    \fi
  }
  \hspace*{\stretch{2}}
  \caption{Circles of convergence of the power series in
    Eq.~\refeq{gbgenerator} for positive $w$.  The corresponding figures
    for negative $w$ are received by mirroring at the vertical axis.
    The thick lines show the discontinuities of the generating
    functions.
  }
  \label{f:circle}
\end{figure}

\begin{figure}
  \hspace*{\stretch{2}}
  \ifnum\thewithlatexpictures=1
    \setlength{\unitlength}{1cm}
    \begin{picture}(6,3.5)(0,0.5)
      \thinlines\small
      \put(0,1){\vector(1,0){6}}
      \put(6,1.1){\makebox(0,0)[br]{$x$}}
      \put(3,1){\vector(0,1){3}}
      \put(2.9,4){\makebox(0,0)[tr]{$\xi$}}
      \put(1.75,3.5){\line(1,0){2.5}}
      \qbezier(5.5,1)(5.5,1.66987)(5.33253,2.25)
      \qbezier(5.33253,2.25)(4.97169,3.5)(4.25,3.5)
      \qbezier(0.5,1)(0.5,1.66987)(0.66747,2.25)
      \qbezier(0.66747,2.25)(1.02831,3.5)(1.75,3.5)
      \dashline[+30]{.1}(3,1)(1.75,3.5)
      \dashline[+30]{.1}(3,1)(4.25,3.5)
      \put(5.5,0.9){\line(0,1){0.2}}
      \put(5.5,0.9){\makebox(0,0)[t]{$+1\vphantom{\tfrac{1}{2}}$}}
      \put(0.5,0.9){\line(0,1){0.2}}
      \put(0.5,0.9){\makebox(0,0)[t]{$-1\vphantom{\tfrac{1}{2}}$}}
      \put(4.25,0.9){\line(0,1){0.2}}
      \put(4.25,0.9){\makebox(0,0)[t]{$+\tfrac{1}{2}$}}
      \put(1.75,0.9){\line(0,1){0.2}}
      \put(1.75,0.9){\makebox(0,0)[t]{$-\tfrac{1}{2}$}}
    \end{picture}
  \else
    \epsfig{figure=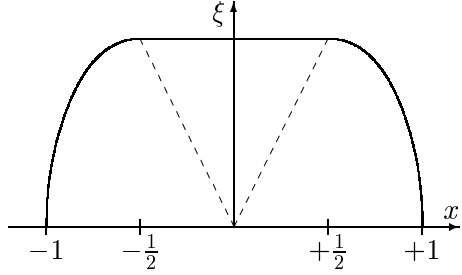}
  \fi
  \hspace*{\stretch{2}}
  \caption{Support area of the effective forward parton distributions.
    The dashed lines correspond to $x=\pm\xi/2$.  The contribution of
    the meson-wave-function-like part of the OFPDs is restricted to the
    region between the dashed lines.} 
  \label{f:support}
\end{figure}

We start with connecting the effective forward parton distribution to
the off-forward ones.  This can be done by Shuvaev's inverse integral
transformation~\cite{Shuvaev:1999fm}:
\begin{subequations}\begin{align}
  \efp{q}{\xi}{x}&=\intoned[{\invkernel{q}{x}{\xi}{y}\ofh{q}{y}{\xi}}]{y
    }{-1}{+1}\mcomma\\
  \efp{g}{\xi}{x}&=\intoned[{\invkernel{g}{x}{\xi}{y}\ofh{g}{y}{\xi}}]{y
    }{-1}{+1}\mperiod
\end{align}\end{subequations}
We briefly sketch the main steps of the derivation of the integral
kernels $\invkernel{q,g}{x}{\xi}{y}$ in Ref.~\cite{Shuvaev:1999fm}, in
order to determine the support properties of the EFPDs. The calculation
is based on the formal inversion of the Mellin moments in
Eq.~\refeq{efpd}
\begin{subequations}\begin{align}
  \efp{q}{\xi}{x}&=-\frac{1}{\pi}\disc\sum_{n=0}^{\infty}\frac{\gbmoment
    {n}{q}{\xi}}{x^{n+1}}\mcomma\\
  \efp{g}{\xi}{x}&=-\frac{1}{\pi}\disc\sum_{n=1}^{\infty}\frac{\gbmoment
    {n}{g}{\xi}}{x^{n+1}}\mcomma\\
  \intertext{with}\disc{}F(x)&=\frac{1}{2i}\lim_{\varepsilon\to0}\left[
    F(x+i\varepsilon)-F(x-i\varepsilon)\right]\mperiod\notag
\end{align}\end{subequations}
The Gegenbauer moments $\gbmoment{n}{q,g}{x}$ are defined in
Eq.~\refeq{gm}.  The factorial functions in Eq.~\refeq{gm} are replaced
using the integral representation of the beta function \{Eqs.~(6.1.18)
and (6.2.1) in Ref.~\cite{Abramowitz:hmf:9}\}.  We obtain
\begin{subequations}\label{eq:preinvkernel}\begin{align}
  \efp{q}{\xi}{x}&=\intoned[{\left[-\frac{1}{\pi}\disc\intoned[{\frac{1}
    {2x\sqrt{1-s}}\sum_{n=0}^{\infty}\gbpolynom{n}{3/2}{\frac{y}
    {\xi}}\left(\frac{s\xi}{2x}\right)^{n}}]{s}{-1}{+1}\right]\ofh{q}{y
    }{\xi}}]{y}{-1}{+1}\mcomma\\
  \efp{g}{\xi}{x}&=\intoned[{\left[-\frac{1}{\pi}\disc\intoned[{\frac{3
    \sqrt{1-s}}{2x^{2}}\sum_{n=0}^{\infty}\gbpolynom{n}{5/2}{
    \frac{y}{\xi}}\left(\frac{s\xi}{2x}\right)^{n}}]{s}{-1}{+1}\right]
    \ofh{g}{y}{\xi}}]{y}{-1}{+1}\mperiod
\end{align}\end{subequations}
The expressions in square brackets are the integral kernels
$\invkernel{q,g}{x}{\xi}{y}$.  Before we state their final form, we have
to look at the generating functions of the Gegenbauer polynomials
\{Eq.~(22.9.3) in Ref.~\cite{Abramowitz:hmf:9}\}:
\begin{equation}\label{eq:gbgenerator}
  \exp\bigl(-\nu\Log(1-2wz+z^{2})\bigr)=\sum_{n=0}^{\infty}\gbpolynom{n}
    {\nu}{w}z^{n}\mperiod
\end{equation}
The generating functions on the left-hand side analytically continue the
power series on the right-hand side to the complete complex plane.  In
Fig.~\ref{f:circle}%
we show the circles of convergence of the power series and the
discontinuities of the generating functions, which arise from negative
arguments of the complex logarithm.  We see that we have to distinguish
two cases: $\abs{y}>\xi$, which corresponds to the
parton-distribution-like region of the OFPDs, and $\abs{y}<\xi$, which
corresponds to the meson-wave-function-like region.  Let us begin with
the latter case.  One might think that for $\abs{y}<\xi$ one has no
contribution to the integral kernels $\invkernel{q,g}{x}{\xi}{y}$
because the generating functions are analytical for any finite and real
$z$. However, the discontinuity at $z=\infty$ produces delta functions
and their derivatives $\opdelta[(n)]{x}$.  From the circle of
convergence we find that their contribution is in any case restricted to
$\abs{x}<\xi/2$ in the effective forward parton distributions.  For
$\abs{y}>\xi$ the discontinuity can be easily calculated.  We face
strong singularities in the generating functions at the end points of
the cut, therefore, we have to take derivatives of less singular
functions, so that the $s$-integral in Eq.~\refeq{preinvkernel} is
convergent.  An examination of the circle of convergence shows that
\begin{equation}\label{eq:support}
  \efp{q}{\xi}{x}=0\text{ and }\efp{g}{\xi}{x}=0\mcomma\quad
    \text{if $\displaystyle\abs{x}>x_{b}\defeq\frac{1}{2}\left(1+\sqrt{1
    -\xi^{2}}\right)$}\mperiod
\end{equation}
This defines the support area of the EFPDs, which is shown in
Fig.~\ref{f:support}.%
Finally, the complete result for the integral kernels of Shuvaev's
inverse transformation is
\begin{subequations}\label{eq:invkernel}\begin{align}
  \invkernel{q}{x}{\xi}{y}&=\frac{2x}{\pi\abs{\xi}}\frac{\partial}{
    \partial{}y}\intoned[{\frac{\optheta{-s^{2}+s\frac{4xy}{\xi^{2}}-
    \frac{4x^{2}}{\xi^{2}}}}{s\sqrt{(1-s)(-s^{2}+s\frac{4xy}{\xi^{2}}-
    \frac{4x^{2}}{\xi^{2}})}}}]{s}{0}{1}\notag\\
  &\phantom{={}}+\optheta{1-\abs{\frac{y}{\xi}}}\sum_{n=0}^{\infty}(-1
    )^{n}\frac{2^{n}[n!]^{2}}{(2n+1)!}\xi^{n}\gbpolynom{n}{3/2
    }{\frac{y}{\xi}}\opdelta[(n)]{x}\mcomma\\
  \invkernel{g}{x}{\xi}{y}&=\frac{4x}{\pi\abs{\xi}}\frac{\partial^{2}}
    {\partial{}y^{2}}\intoned[{\frac{\optheta{-s^{2}+s\frac{4xy}{\xi^{
    2}}-\frac{4x^{2}}{\xi^{2}}}\sqrt{1-s}}{s^{2}\sqrt{-s^{2}+s\frac{
    4xy}{\xi^{2}}-\frac{4x^{2}}{\xi^{2}}}}}]{s}{0}{1}\notag\\
  &\phantom{={}}+\optheta{1-\abs{\frac{y}{\xi}}}\sum_{n=1}^{\infty}(-1
    )^{n}\frac{3\cdot2^{n}(n-1)!n!}{(2n+1)!}\xi^{n-1}\gbpolynom{n-1}{
    5/2}{\frac{y}{\xi}}\opdelta[(n)]{x}\mperiod
\end{align}\end{subequations}
We cannot carry out the derivatives, which would lead to infinite
contributions from the derivatives of the theta functions and divergent
integrals. The infinite sums represent the contribution of the
meson-wave-function-like part of the off-forward parton distributions
and do not appear in Ref.~\cite{Shuvaev:1999fm}, since the discontinuity
at $z=\infty$, resp.\ $x=0$, was overlooked.  The occurence of delta
functions and their derivatives in the effective forward parton
distributions are comparable to meson-exchange-type contributions in
double distributions~\cite{Radyushkin:1998bz,Musatov:1999xp}.

The expressions for the integral kernels $\invkernel{q,g}{x}{\xi}{y}$
show that the Gegenbauer moment inversion is not practicable, in
general.  Therefore, it can generally not be used for a simple solution
of the evolution equations of off-forward parton distributions.
%
%
\section{Shuvaev's transformation}
\label{s:kernel}
As we will later see, the predictive power of the formalism lies in
relating the effective forward parton distributions to the off-forward
ones by Shuvaev's integral transformation
\begin{subequations}\label{eq:trafo}\begin{align}
  \ofh{q}{x}{\xi}&=\intoned[{\kernel{q}{x}{\xi}{y}\efp{q}{\xi}{y}}]{y}{-
    1}{+1}\mcomma\\
  \ofh{g}{x}{\xi}&=\intoned[{\kernel{g}{x}{\xi}{y}\efp{g}{\xi}{y}}]{y}{-
    1}{+1}\mperiod
\end{align}\end{subequations}
The full derivation of the integral kernels $\kernel{q,g}{x}{\xi}{y}$
can be found in Ref.~\cite{Shuvaev:1999fm}.  We merely state the finite
result:
\begin{subequations}\label{eq:kernel}\begin{align}
  \kernel{q}{x}{\xi}{y}&=\frac{1}{\pi\sqrt{\abs{y}}}\frac{\partial}{
    \partial{}y}\frac{y}{\sqrt{\abs{y}}}\intoned[{\optheta{\frac{y(1-s^{
    2})}{x-\xi{}s}-1}\sqrt{\frac{x-\xi{}s}{y(1-s^{2})-x+\xi{}s}}}]{s}{-1
    }{+1}\mcomma\\
  \kernel{g}{x}{\xi}{y}&=\frac{1}{\pi\sqrt{\abs{y}}}\frac{\partial}{
    \partial{}y}\sqrt{\abs{y}}\intoned[{\optheta{\frac{y(1-s^{2})}{x-\xi
    {}s}-1}\sqrt{\frac{(x-\xi{}s)^{3}}{y(1-s^{2})-x+\xi{}s}}}]{s}{-1}{+1
    }\mperiod
\end{align}\end{subequations}
Again, performing the derivatives would give divergent integrals and
infinite contributions from the end points.  The derivatives of the
theta function give rise to the ``suspicious overall sign'' that is
mentioned in Ref.~\cite{Musatov:1999xp}.  Equation~\refeq{kernel} is
equivalent to previous results presented in
Refs.~\cite{Shuvaev:1999ce,Musatov:1999xp}.

It is useful to express the integral kernels in terms of standard
elliptic integrals, because it is then possible to perform the
derivatives analytically. First, we give the symmetry properties of the
integral kernels:
\begin{subequations}\label{eq:symmetry}\begin{align}
  \kernel{q}{x}{-\xi}{y}&=\kernel{q}{x}{\xi}{y}\mcomma&\kernel{q}{x}{\xi
    }{-y}&=+\kernel{q}{-x}{\xi}{y}\mcomma\\
  \kernel{g}{x}{-\xi}{y}&=\kernel{g}{x}{\xi}{y}\mcomma&\kernel{g}{x}{\xi
    }{-y}&=-\kernel{g}{-x}{\xi}{y}\mperiod
\end{align}\end{subequations}
Of course, $\kernel{q,g}{x}{\xi}{y}$ obey the fundamental
$(-\xi\leftrightarrow\xi)$-symmetry of Ji's off-forward parton
distributions.  The relations on the right-hand side show that it is
sufficient to restrict the calculation to positive values of $y$.  We
define
\begin{equation}
  a\defeq\frac{x}{\xi}-\frac{\xi}{2y}-\frac{1}{2y}\sqrt{4y^{2}-4yx+\xi^{
    2}}\mcomma\quad{}b\defeq\frac{x}{\xi}-\frac{\xi}{2y}+\frac{1}{2y}
    \sqrt{4y^{2}-4yx+\xi^{2}}\mcomma
\end{equation}
which correspond up to the $x/\xi$, which we added for convenience, to
the zeroes of the denominator in the square root in Eq.~\refeq{kernel}.
As the radicand has to be positive, we must further restrict the
possible range of $y$ to
\begin{equation}
  y>x_{a}\defeq\frac{1}{2}(x+\sqrt{x^{2}-\xi^{2}})\mcomma\quad\text{if $
    x\geq\xi\mperiod$}
\end{equation}
With help of the integral tables in Ref.~\cite{Carlson:1989} we obtain
our final result:
\begin{subequations}\label{eq:mykernel}\begin{align}
  \kernel{q}{x}{\xi}{y}&=
  \begin{cases}
    \opdelta{y-x_{a}}\sqrt{\frac{1}{x_{a}}\sqrt{x^{2}-\xi^{2}}}&\\
    \quad+\optheta{y-x_{a}}\frac{\xi}{\pi{}y^{2}}\sqrt{\frac{\xi}{yb}
      }\frac{b}{b-a}\left(\vphantom{\frac{-a}{b-a}}\carlson{F}{0}{\frac{
      a}{b}}{1}-\frac{1}{3}\frac{b+a}{b}\carlson{D}{0}{\frac{a}{b}}{1}
      \right)&\text{for $x\ge\xi$}\mcomma\\
    \optheta{x+\xi}\frac{\xi}{\pi{}y^{2}}\sqrt{\frac{\xi}{y(b-a)}}\frac{
      b}{b-a}\left(\carlson{F}{0}{\frac{-a}{b-a}}{1}-\frac{1}{3}\frac{b+
      a}{b-a}\carlson{D}{0}{\frac{-a}{b-a}}{1}\right)&\text{for $x<\xi$}
      \mcomma
  \end{cases}\\
  \kernel{g}{x}{\xi}{y}&=
  \begin{cases}
    \opdelta{y-x_{a}}\sqrt{\frac{1}{x_{a}}(x^{2}-\xi^{2})^{3/2}}+
      \optheta{y-x_{a}}\frac{\xi^{2}}{\pi{}y^{2}}\sqrt{\frac{\xi{}b}{y
      }}\frac{b}{b-a}&\\
    \quad\times\left(\frac{2b-a}{b}\carlson{F}{0}{\frac{a}{b}}{1}-
      \frac{2}{3}\frac{b^{2}-ab+a^{2}}{b^{2}}\carlson{D}{0}{\frac{a}{b
      }}{1}\right)&\text{for $x\ge\xi$}\mcomma\\
    \optheta{x+\xi}\frac{\xi^{2}}{\pi{}y^{2}}\sqrt{\frac{\xi(b-a)}{y}}
      \frac{b}{b-a}&\\
    \quad\times\left(\frac{2b-a}{b-a}\carlson{F}{0}{\frac{-a}{b-a}}{1
      }-\frac{2}{3}\frac{b^{2}-ab+a^{2}}{(b-a)^{2}}\carlson{D}{0}{
      \frac{-a}{b-a}}{1}\right)&\text{for $x<\xi$}\mcomma
  \end{cases}
\end{align}\end{subequations}
where $R_{F}$ and $R_{D}$ are Carlson's elliptic integrals of the first
and second kind (see, e.g., \cite{Carlson:1979}) with
\begin{align}
  \carlson{F}{x}{y}{z}&\defeq\frac{1}{2}\intoned[{\frac{1}{\sqrt{(t+x)(t
    +y)(t+z)}}}]{t}{0}{\infty}\mcomma\\
  \carlson{D}{x}{y}{z}&\defeq\frac{3}{2}\intoned[{\frac{1}{(t+z)\sqrt{(t
    +x)(t+y)(t+z)}}}]{t}{0}{\infty}\mperiod
\end{align}
These integral kernels accumulate the main properties of the $x$ and
$\xi$ dependence of off-forward parton distributions.  For $x\gg\xi$,
the OFPDs essentially look like forward quarks and gluons:
\begin{subequations}\begin{align}
  \kernel{q}{x}{\xi}{y}&=\opdelta{y-x}+\frac{1}{x}\landauO{\frac{\xi^{2}
    }{x^{2}}}\mcomma\\
  \kernel{g}{x}{\xi}{y}&=x\opdelta{y-x}+\landauO{\frac{\xi^{2}}{x^{2}}}
    \mperiod
\end{align}\end{subequations}

The forward evolution concentrates the effective forward parton
distributions at $y\sim0$.  Therefore, the small-$y$ behavior of the
integral kernels $\kernel{q,g}{x}{\xi}{y}$ reproduces the well-known
asymptotic forms of the off-forward valence and singlet quark, and gluon
distributions~\cite{Radyushkin:1997ki}:
\begin{subequations}\begin{align}
  \begin{split}
  \ofh[v]{q}{x}{\xi}&=\optheta{1-\abs{\frac{x}{\xi}}}\frac{\xi^{2}-x^{2
    }}{\xi^{3}}\intoned[{\left(\frac{3}{2}+\landauO{\frac{y}{\xi}}\right
    )\efp[v]{q}{\xi}{y}}]{y}{0}{\varepsilon}\mcomma\\
  \ofh[s]{q}{x}{\xi}&=\optheta{1-\abs{\frac{x}{\xi}}}\frac{x\left(\xi^{2
    }-x^{2}\right)}{\xi^{5}}\intoned[{\left(\frac{15}{2}+\landauO{\frac{
    y}{\xi}}\right)y\efp[s]{q}{\xi}{y}}]{y}{0}{\varepsilon}\mcomma
  \end{split}\\
  \ofh{g}{x}{\xi}&=\optheta{1-\abs{\frac{x}{\xi}}}\frac{(\xi^{2}-x^{2})^
    {2}}{\xi^{5}}\intoned[{\left(\frac{15}{8}+\landauO{\frac{y}{\xi}}
    \right)y\efp{g}{\xi}{y}}]{y}{0}{\varepsilon}\mperiod
\end{align}\end{subequations}
Additionally, these equations prove that the integrals in
Eq.~\refeq{trafo} are well defined and convergent.

Even the physical interpretation of the different regions in $x$ holds
in the formalism of EFPDs. Inserting Eq.~\refeq{symmetry} in
Eq.~\refeq{trafo} yields
\begin{subequations}\begin{align}
  \ofh{q}{x}{\xi}&=\intoned[{\bigl(\kernel{q}{x}{\xi}{y}\efp{q}{\xi}{y}+
    \kernel{q}{-x}{\xi}{y}\efp{q}{\xi}{-y}\bigr)}]{y}{0}{1}\mcomma\\
  \ofh{g}{x}{\xi}&=\intoned[{\bigl(\kernel{g}{x}{\xi}{y}+\kernel{g}{-x}{
    \xi}{y}\bigr)\efp{g}{\xi}{y}}]{y}{0}{1}\mperiod
\end{align}\end{subequations}
For $\abs{x}>\xi$ the off-forward parton distributions are related to
corresponding effective forward (anti)partons with a minimum momentum
$\abs{x_{a}}$.  For $\abs{x}<\xi$ the picture of a meson wave function
is supported by a simultaneous contribution of effective forward partons
and antipartons with any momentum $y$.  The different expressions for
the integral kernels $\kernel{q,g}{x}{\xi}{y}$ for $\abs{x}\gtrless\xi$
show, analogous to~\cite{Radyushkin:1998bz}, that the off-forward parton
distributions are not analytic at $\abs{x}=\xi$.  The analyticity of the
OFPDs for $x\neq\xi$ requires that the effective forward parton
distributions need to be analytic for $\abs{x}\ge\xi/2$ only.
%
%
\section{Nonforward amplitudes and EFPDs}
\label{s:amplitude}
The use of off-forward parton distributions is required in deeply
virtual Compton scattering (DVCS) and hard exclusive electroproduction
processes.  Detailed information can be found in Ref.~\cite{Ji:1998pc},
and references therein.  Here, we are only interested in the part of the
amplitudes that refers to the OFPDs:
\begin{subequations}\label{eq:amplitude}\begin{align}
  \amplitude{q}&\defeq\intoned[{\left(\frac{1}{x-\xi+i\varepsilon}+\frac
    {1}{x+\xi-i\varepsilon}\right)\ofh{q}{x}{\xi}}]{x}{-1}{+1}\mcomma\\
  \amplitude{g}&\defeq\intoned[{\left(\frac{1}{x-\xi+i\varepsilon}+\frac
    {1}{x+\xi-i\varepsilon}\right)\frac{1}{x}\ofh{g}{x}{\xi}}]{x}{-1}{+1
    }\mcomma
\end{align}\end{subequations}
where we have neglected the $\landauO{\Delta}$ contributions, analogous
to Ref.~\cite{Shuvaev:1999ce}.  The imaginary part of the amplitudes is
related to the diagonal elements $\ofh{q,g}{\xi}{\xi}$, which can be
expressed by the effective forward parton distributions:
\begin{subequations}\label{eq:imagpart}\begin{alignat}{2}
  -\pi\ofh[s]{q}{\xi}{\xi}&=\imagpart\amplitude{q}&&=-\intoned[{4\sqrt{1
    -z^{2}}\efp[s]{q}{\xi}{\tfrac{\xi}{2(1-z^{2})}}}]{z}{0}{\sqrt{1-
    \xi/2x_{b}}}\mcomma\\
  -\frac{2\pi}{\xi}\ofh{g}{\xi}{\xi}&=\imagpart\amplitude{g}&&=-\intoned
    [{32z^{2}\sqrt{1-z^{2}}\efp{g}{\xi}{\tfrac{\xi}{2(1-z^{2})}}}]{z}{0}
    {\sqrt{1-\xi/2x_{b}}}\mcomma
\end{alignat}\end{subequations}
with the quark singlet
$\efp[s]{q}{\xi}{x}=\efp{q}{\xi}{x}-\efp{q}{\xi}{-x}$.  The imaginary
part is essentially dominated by the behavior of the EFPDs around
$x\sim\xi/2$.  For small values of $\xi$ this region can be accurately
described by
\begin{subequations}\label{eq:lambda}\begin{align}
  x\efp[s]{q}{\xi}{x}\sim{}x^{-\lambda_{q}}\mcomma\\
  x\efp{g}{\xi}{x}\sim{}x^{-\lambda_{g}}\mperiod
\end{align}\end{subequations}
If we insert Eq.~\refeq{lambda} into Eq.~\refeq{imagpart} and set the
upper integration limits to one, we achieve approximation formulas for
the imaginary part of the amplitudes in Eq.~\refeq{amplitude}:
\begin{subequations}\label{eq:im2efp}\begin{align}
  R_{q}^{\imagpart}\defeq\frac{\imagpart\amplitude{q}}{\efp[s]{q}{\xi}{
    \tfrac{\xi}{2}}}&\simeq-\frac{2\sqrt{\pi}\opGamma{\lambda_{q}+\tfrac
    {5}{2}}}{\opGamma{\lambda_{q}+3}}\mcomma\\
  R_{g}^{\imagpart}\defeq\frac{\imagpart\amplitude{g}}{\efp{g}{\xi}{
    \tfrac{\xi}{2}}}&\simeq-\frac{8\sqrt{\pi}\opGamma{\lambda_{g}+\tfrac
    {5}{2}}}{\opGamma{\lambda_{g}+4}}\mperiod
\end{align}\end{subequations}
A similar ratio was already presented in Ref.~\cite{Shuvaev:1999ce},
where, however, the imaginary part was compared to diagonal partons at
$x=2\xi$, which leads to an extra factor $2^{2+2\lambda_{q,g}}$.

\begin{figure}
  \hspace*{\stretch{2}}
  \subfigure[$\displaystyle{}R_{q}^{\imagpart}(\text{exact}):R_{q}^{
    \imagpart}(\lambda_{q})$]{\epsfig{figure=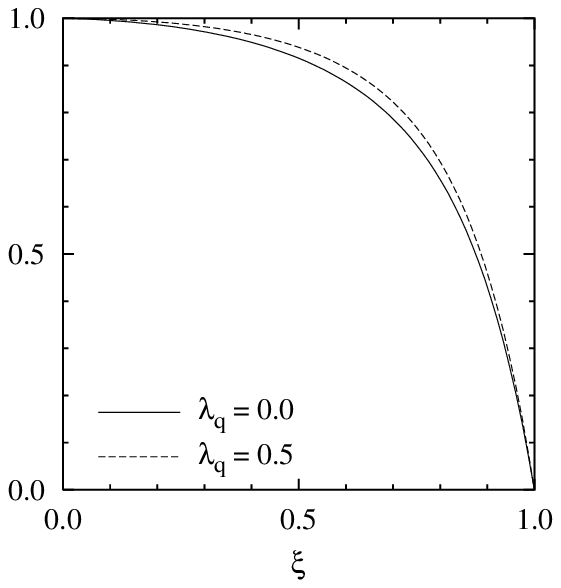}}
  \hspace*{\stretch{1}}
  \subfigure[$\displaystyle{}R_{g}^{\imagpart}(\text{exact}):R_{g}^{
    \imagpart}(\lambda_{g})$]{\epsfig{figure=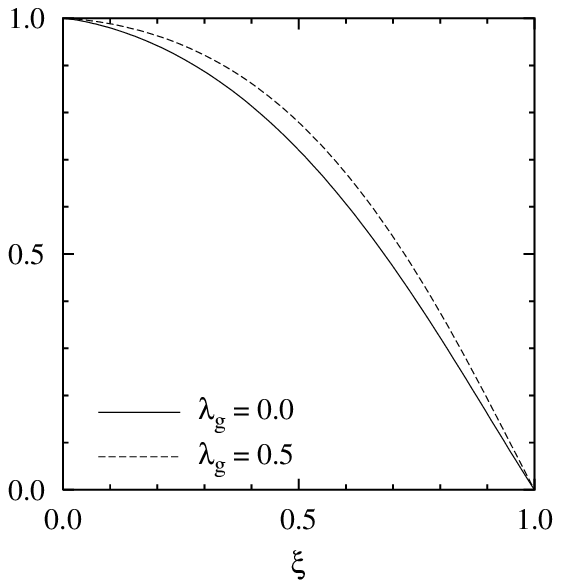}}
  \hspace*{\stretch{2}}
  \caption{Comparison of the exact imaginary part to the approximative
    ratio $R_{q,g}^{\imagpart}(\lambda_{q,g})$ for $\efp[s]{q}{\xi}{x},$
    $\efp{g}{\xi}{x}\sim{}x^{-\lambda_{q,g}-1}\mperiod$}
  \label{f:im2efp}
\end{figure}

In Fig.~\ref{f:im2efp}%
we show a comparison of the exact ratio, derived from
Eq.~\refeq{imagpart}, to the approximation in Eq.~\refeq{im2efp} for the
effective distributions in Eq.~\refeq{lambda}.  The change of the
integration limit has no remarkable effect up to $\xi\sim0.1$.  The
accuracy of the gluon ratio is slightly worse compared to the quark
ratio, because of an additional factor of $z^{2}$ in the integrand in
Eq.~\refeq{imagpart}.  Since the quotients of the gamma functions in
Eq.~\refeq{im2efp} have a weak $\lambda_{q,g}$ dependency and
Fig.~\ref{f:im2efp} shows a good stability under a change of
$\lambda_{q,g}$, we can conclude that Eq.~\refeq{im2efp} is an excellent
approximation of the imaginary part of the amplitudes for the values of
$\xi$, where the effective forward parton distributions can be reliably
identified with the conventional forward quark and gluon densities.

\begin{figure}
  \hspace*{\stretch{2}}
  \subfigure[$\displaystyle{}R_{q}^{\realpart}(\lambda_{q}):R_{q}^{
    \realpart}(\text{exact})$]{\epsfig{figure=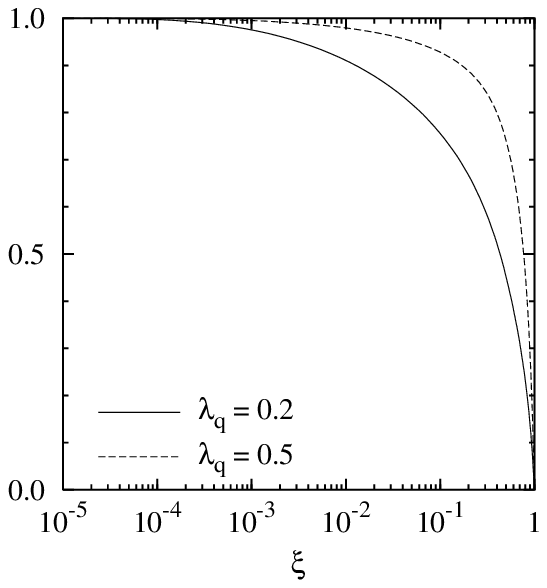}}
  \hspace*{\stretch{1}}
  \subfigure[$\displaystyle{}R_{g}^{\realpart}(\lambda_{g}):R_{g}^{
    \realpart}(\text{exact})$]{\epsfig{figure=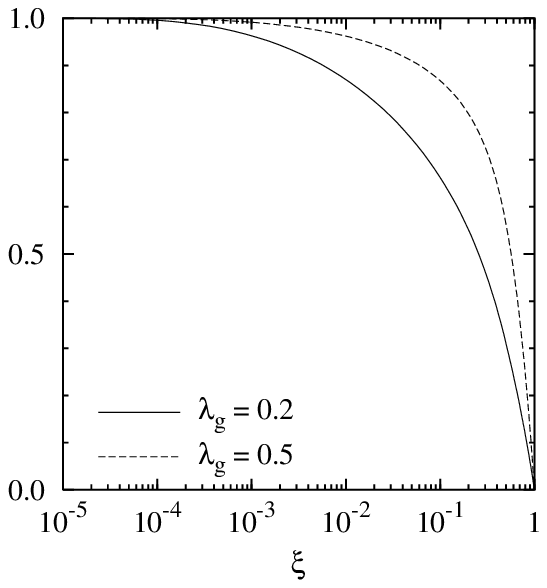}}
  \hspace*{\stretch{2}}
  \caption{Comparison of the exact real part to the approximative ratio
    $R_{q,g}^{\realpart}(\lambda_{q,g})$ for $\efp[s]{q}{\xi}{x},$
    $\efp{g}{\xi}{x}\sim{}x^{-\lambda_{q,g}-1}\mperiod$}
  \label{f:re2im}
\end{figure}

The calculation of the real part is straightforward but tedious [we have
used Eq.~\refeq{kernel} rather than Eq.~\refeq{mykernel}].  The
principal value integration can be performed exactly and the final
result consists of integrals without any strong singularities:
\begin{subequations}\label{eq:realpart}\begin{align}
  \realpart\amplitude{q}&=\intoned[{\frac{2}{z^{2}}\left(z+\frac{1}{
    \sqrt{1+z}}-\frac{1}{\sqrt{1-z}}\right)\efp[s]{q}{\xi}{\tfrac{\xi{
    }z}{2}}}]{z}{0}{1}\notag\\
  &\phantom{={}}+\intoned[{2\left(\frac{1}{z}+\sqrt{\frac{z}{1+z}}
    \right)\efp[s]{q}{\xi}{\tfrac{\xi}{2z}}}]{z}{\xi/2x_{b}}{1}\mcomma\\
  \realpart\amplitude{g}&=\intoned[{\frac{4}{z^{3}}\left(z^{2}-8+4
    \sqrt{1+z}+4\sqrt{1-z}\vphantom{\frac{1}{\sqrt{1-z}}}\right)\efp{g
    }{\xi}{\tfrac{\xi{}z}{2}}}]{z}{0}{1}\notag\\
  &\phantom{={}}+\intoned[{4\left(\frac{1}{z}-8z+4\sqrt{z(1+z)}\right)
    \efp{g}{\xi}{\tfrac{\xi}{2z}}}]{z}{\xi/2x_{b}}{1}\mperiod
\end{align}\end{subequations}
We note that the expressions in brackets in the first integrals in
Eq.~\refeq{realpart} are always negative, therefore, the two integrals
partly cancel each other.  Again, we insert the small-$x$ behavior of
Eq.~\refeq{lambda} into these integrals and set the lower limits of the
second integrals to zero, so that everything can be evaluated and yields
the following ratios between the real and imaginary parts:
\begin{equation}\label{eq:re2im}
  R_{q,g}^{\realpart}\defeq\frac{\realpart\amplitude{q,g}}{\imagpart
    \amplitude{q,g}}\simeq\tan\frac{\pi\lambda_{q,g}}{2}\mperiod
\end{equation}
This is identical to the result, achieved by dispersion relations, in
Refs.~\cite{Frankfurt:1998at,Shuvaev:1999ce}. From Fig.~\ref{f:re2im}%
we see that the quality of the approximations of the real parts is
significantly more sensitive --- note the logarithmic scale --- to the
change of the integration limit and to a variation of $\lambda_{q,g}$.
Additionally, the right-hand side of Eq.~\refeq{re2im} depends strongly
on $\lambda_{q,g}$, and the first integrals in Eq.~\refeq{realpart} have
dominant contributions from two regions: around $x\sim0$ and
$x\sim\xi/2$, i.e., we must require that $\lambda_{q,g}$ is essentially
constant for small $x$.  Therefore, only for very small $\xi$, when
Eq.~\refeq{lambda} is a valid approximation for a large range of $x$ for
the usual forward quark and gluon distribution, the latter can be used
to reliably predict the real part of the amplitude.  Nevertheless, for
small values of $\lambda_{q,g}$, where the real part is strongly
suppressed, the absolute values of the amplitudes are determined to a
good precision for small $\xi$~\cite{Shuvaev:1999ce}.
%
%
\section{Moment-diagonal models}
\label{s:model}
In this section, we try to build a model for the effective forward
parton distributions.  The situation does not seem to be very promising,
because Shuvaev's inverse transformation cannot generally be used and we
face a difficult support area for the EFPDs in Fig.~\ref{f:support}.

References~\cite{Shuvaev:1999ce,Golec-Biernat:1999ib,Musatov:1999xp}
were dealing with a model for off-forward parton distributions, the
Gegenbauer moments of which are independent of $\xi$.  Because the
corresponding EFPDs are also independent of $\xi$, these models
manifestly violate the support area in Fig.~\ref{f:support}, as was also
recognized in Ref.~\cite{Musatov:1999xp}, though it should be a good
approximation for small values of $\xi$. This model would have had the
great advantage that it would have been stable against a change of the
input scale, since the evolution of the Gegenbauer moments is identical
to that of the Mellin moments.

Nevertheless, this idea can be used to find valid models for the
effective forward parton distributions.  Because a common $n$-, $\xi$-,
or $t$-dependent factor does not have any influence on the evolution
equations, we can generalize the model with $\xi$-independent Gegenbauer
moments to a class of moment-diagonal models with a common
proportionality factor:
\begin{equation}\label{eq:family}
  \gbmoment{n}{q,g}{\xi,t,\mu}\defeq\text{const}(n,\xi,t)\times\mmoment{
    n}{q,g}(\mu)\mperiod
\end{equation}
The $t$ dependence is usually factorized,
$\text{const}(n,\xi)\times{}F_{1}(t)$, where $F_{1}(t)$ is the Dirac
form factor~\cite{Shuvaev:1999ce,Golec-Biernat:1999ib}.  These models
allow, provided that the Mellin inverse can be performed and leads to
valid supports, a direct and simple calculation of off-forward parton
distributions and amplitudes, with help of the formulas
\refeq{mykernel}, \refeq{imagpart}, and \refeq{realpart}, for arbitrary
$x$, $\xi$, $\mu$, and for the region of $t$, where the
$\landauO{\Delta}$ contributions can be neglected.  The simplest model
for proton OFPDs of this class, which fulfills all known theoretical
constraints for $t=0$ (e.g., see Ref.~\cite{Ji:1998pc}) and gives a good
approximation of the $t$-dependence, is
\begin{equation}
  \gbmoment{n}{q,g}{\xi,t,\mu}\defeq2\left(\frac{\xi}{2}\right)^{n+1}
    \chpolynom{n+1}{\xi^{-1}}\mathop{{}F^{p}_{1}(t)}\mmoment{n}{q,g}(\mu
    )\mcomma
\end{equation}
where $\mmoment{n}{q,g}(\mu)$ are the Mellin moments of the usual quark
and gluon distributions in the proton, $F_{1}^{p}(t)$ is the Dirac form
factor of the proton with $F_{1}^{p}(0)=1$, and $T_{n}(x)$ are Chebyshev
polynomials of the first kind \{Eq.~(22.2.4) in
Ref.~\cite{Abramowitz:hmf:9}\}.  It is advantageous to use the
Gl{\"{u}}ck--Reya--Vogt 1998 (GRV~98) parton
distributions~\cite{Gluck:1998xa}, since they are given for $x$ values
down to $10^{-9}$, which allows us to compute the real part in
Eq.~\refeq{realpart} for small values of $\xi$ with a high accuracy.
With use of Eq.~(22.3.25) and (4.4.27) in Ref.~\cite{Abramowitz:hmf:9},
a Mellin inversion yields
\begin{subequations}\label{eq:model}\begin{align}
  \efp{q}{\xi}{x}&=\left\{\optheta{\tfrac{1+\sqrt{1-\xi^{2}}}{2}-\abs{x}
    }\efp{q}{}{\tfrac{2x}{1+\sqrt{1-\xi^{2}}}}+\optheta{\tfrac{1-\sqrt{1
    -\xi^{2}}}{2}-\abs{x}}\efp{q}{}{\tfrac{2x}{1-\sqrt{1-\xi^{2}}}}
    \right\}\mathop{F_{1}^{p}(t)}\mcomma\\
  \efp{g}{\xi}{x}&=\left\{\optheta{\tfrac{1+\sqrt{1-\xi^{2}}}{2}-\abs{x}
    }\efp{g}{}{\tfrac{2x}{1+\sqrt{1-\xi^{2}}}}+\optheta{\tfrac{1-\sqrt{1
    -\xi^{2}}}{2}-\abs{x}}\efp{g}{}{\tfrac{2x}{1-\sqrt{1-\xi^{2}}}}
    \right\}\mathop{F_{1}^{p}(t)}\mperiod
\end{align}\end{subequations}
We see that the effective forward parton distributions are a simple
combination of two rescaled forward parton densities with the correct
support area and an appropriate common $t$-dependent factor.  The first
one gives the conventional forward quark and gluons for vanishing $\xi$
and $t$. The contribution of the second summand is restricted to
$\abs{x}<\xi/2$, i.e., it influences only the meson-wave-function-like
region $\abs{x}<\xi$ of OFPDs, and is negligible for small values of
$\xi$.  Therefore, most of the numerical results of
Refs.~\cite{Shuvaev:1999ce,Golec-Biernat:1999ib,Musatov:1999xp} can be
accurately transferred to the model in Eq.~\refeq{model}.

It is an interesting fact that the argument of the first forward parton
density in Eq.~\refeq{model} is very similar to the Georgi--Politzer
$\xi$-scaling variable~\cite{Georgi:1976ve},
$\xi=2x_{B}/(1+\sqrt{1+4x_{B}^{2}M_{N}^{2}/Q^{2}})$, that originally
described target mass effects in deep inelastic scattering.  Hence, one
can argue that the arguments of the parton densities in
Eq.~\refeq{model} reflect skewedness effects.  But such arguments can as
well be relics of the Gegenbauer polynomials that appear in the
derivation~\cite{Nachtmann:1973mr,Barbieri:1976rd} of the $\xi$-scaling
variable.
%
%
\section{Summary and conclusions}
\label{s:summary}
In this paper, we presented with Eqs.~\refeq{mykernel},
\refeq{imagpart}, and \refeq{realpart} simple expressions that relate
effective forward parton distributions, which evolve like conventional
forward partons, to off-forward parton distributions and nonforward
amplitudes.  We emphasized that the off-forward parton distributions and
nonforward amplitudes can be directly determined from the conventional
forward parton distributions and nucleon form factors at arbitrary scale
$\mu$ for moment-diagonal models.  Exemplary, we stated a simple
self-consistent model for the EFPDs of the proton in terms of the GRV~98
parton distributions~\cite{Gluck:1998xa} and the Dirac form factor of
the proton, which allows us to predict off-forward parton distributions
and nonforward amplitudes for arbitrary $x$, $\xi$, $\mu$, and (not to
large) $t$.  These predictions should not differ too much from results
of other models at least at small $\xi$, as the results in
Refs.~\cite{Shuvaev:1999ce,Golec-Biernat:1999ib} show.

Nevertheless, it would be illuminating if further self-consistent
moment-diagonal models exist, especially models that have a
qualitatively different behavior in the meson-wave-function-like region,
such as the off-forward parton distributions of chiral soliton model
calculations~\cite{Petrov:1998kf}, since the real part of nonforward
amplitudes is dominated by this region and gets important for large
$\xi$.

Because of the complicated support area of EFPDs, an investigation of
the double distributions of moment-diagonal models might be helpful.
%
%
\section*{Acknowledgments}
We thank M.~Gl\"{u}ck and E.~Reya for proposing this investigation and
for instructive remarks during the initial stage of this work.
%
%

%
\end{document}